\documentclass[12pt]{article}
\usepackage{amsmath,amssymb}
\usepackage{graphicx, epsf}
\title{Goos-H\"anchen shift for higher order Hermite-Gaussian beams}
\author{Dheeraj Golla$^1$, S Dutta Gupta$^2$ \\
\small{$^1$ Indian Institute of Technology Kharagpur, Kharagpur 721302 , India} \\
\small{$^2$ School of Physics, University of Hyderabad, Hyderabad 500046, India}}
\date{}
\begin{document}


\maketitle
\abstract{
We study the reflection of a Hermite-Gaussian beam at an interface between two dielectric media.  We show that unlike Laguerre-Gaussian beams, Hermite-Gaussian beams undergo no significant distortion upon reflection. We report Goos-H\"anchen shift for all the spots of a higher order Hermite-Gaussian beam near the critical angle. The shift is shown to be insignificant away from the critical angle. The calculations are carried out neglecting the longitudinal component along the direction of propagation for a spatially finite, s-polarized, full 3-d vector beam. We briefly  discuss the difficulties associated with the paraxial approximation pertaining to a vector gaussian beam.}


\section{Introduction}
The Goos-H\"anchen (GH) shift, first discovered experimentally in 1947, refers to the longitudinal displacement of a beam from its expected geometric position under total internal reflection \cite{Goos}.  Artmann \cite{Artmann} was the first to provide a theoretical explanation of the effect using the stationary phase approximation. A somewhat different calculation of the Goos-H\"anchen shift using the conservation of energy was put forth by Renard \cite{Renard}. Quantitatively the GH shift can be  interpreted as the shift of the peak intensity spot of the beam \cite{Artmann} or the weighted mean position of the beam \cite{Woerdman}. A thorough theoretical analysis of the effect was carried out by Lotsch \cite{Lotsch} in 1968. The effect, although discovered quite a while ago, still remains a topic of interest in current optics literature. The current interest in GH shift is further fueled by several means for enhancing the effect and its use for  sensing \cite{Xiaobo}. These include structures involving absorbing media \cite{Wild,Yan}, metals \cite{Merano}, left-handed metamaterials \cite{Shadrivov,Kui}. Very recent studies focus on systems supporting surface plasmon resonances \cite{Xiaobo,Yin,Lin} and frustrated total internal reflection \cite{Haibel}. From a theoretical angle, most of the studies on GH shift used to suffer from two major drawbacks. First, an overwhelming majority deal with two dimensional fundamental beams (the third dimension is supressed) \cite{McGuirk,Chiu,Simon}, and second, they overlook the vector characteristics of the beam. Recall that the paraxial electromagnetic wave is not an exact solution of the Maxwell's equations, but holds only to the first order in $\frac{1}{k}$ ($k$ being the magnitude of the wavevector). Since the pioneering work by Agrawal and Pattanayak \cite{Agrawal}, the difficulties associated with the paraxial approximation  are now clear. The vector nature of the beams adds to the complexity of the problem (leading to a longitudinal field component). Only recently vector Gaussian beams have been dealt with in the context of propagation in a uniform medium  \cite{Carlchen}. A systematic treatment of GH shift for a fundamental vector Gaussian beam was given by Aiello and Woerdman  \cite{Aiello}. It was shown by the same authors in a later paper that the suppression of the correction due to the longitudinal component recovers Artman formula for the shift \cite{Woerdman}. There has been a great deal of interest in the Laguerre-Gaussian beams and their longitudinal and lateral shifts in view of the angular momentum carried by the beam \cite{Barnette,Allen,Simpson,Bliokh,Merano2,Hiroshi}. The longitudinal and transverse shifts for the Laguerre-Gaussian beams were calculated by Bliokh {\it et al} \cite{Bliokh} and verified experimentally by Merano {\it et al} \cite{Merano2}. Another  recent paper presents both theoretical and experimental results on the shift of Laguerre-Gaussian beams and its distortion under reflection and transmission \cite{Hiroshi}. However,  no such studies exist for simpler Hermite-Gaussian beams. In this paper we concentrate on a general vector Hermite-Gaussian beam with s-polarization and look at the reflection of the same from a denser to lighter medium interface. We show how the longitudinal component arises from a simple description of the vector field by means of scalar and vector potentials. Neglecting the longitudinal component under the strong paraxial approximation, we use a angular spectrum decomposition to calculate the reflected field for each spatial harmonic. Summing up the harmonics we end up with the reflected beam profile.
\par 
The structure of the paper is as follows. In section 2, we provide a simple mathematical formulation of the problem and show that a longitudinal component is inevitable in the paraxial formulation. In section 3, we first validate our code by matching our numerical results with those predicted by Artman formula. Next, we present the contour plots for fundamental and higher order Hermite-Gaussian beam profiles. We also comment on the possibility of splitting of the reflected beam, which has been observed for tightly focussed beams. Finally, we summarize the main results in the conclusion.
\section{Mathematical formulation}
Consider the electromagnetic wave of frequency $\omega$ described by the electric $\mathbf{E}$ and magnetic $\mathbf{B}$ fields in a medium with dielectric constant $\epsilon$ and permeability constant $\mu = 1$. Let the scalar and vector potentials be $\phi$ and $\mathbf{A}$, so that
\begin{equation} {\mathbf B} = \nabla\times{{\mathbf A}} = {\mathbf H}, \label{eq:1}\end{equation}
\begin{equation} {\mathbf E} = -\nabla\phi - \frac{1}{c}\frac{\partial{\mathbf A}}{\partial t} = -\nabla\phi + ik_{0}{\mathbf A}, \label{eq:2}\end{equation}
where $ k_{0} = \frac{\omega}{c} $ is the vacuum wave vector.
Under the Lorentz condition, the vector $\mathbf{A}$ satisfies the following equations:
\begin{equation}
\nabla\cdot{{\mathbf A}} - ik_{0}\epsilon\phi = 0, 
\label{eq:3}
\end{equation}
\begin{equation} 
\nabla^{2}{\mathbf A} + k_{0}^{2}\epsilon{\mathbf A} = 0. 
\label{eq:4}
\end{equation} 
One way to tackle the problem is to ignore the variation of all quantities in one direction, making it effectively a two dimensional problem. Polarization is no longer an issue in the two dimensional scenario and the standard scalar approximation techniques can be used. A more general set of solutions is obtained by assuming that the vector ${\mathbf A}$, rather than ${\mathbf E}$ is polarized in a particular direction.
\par
We seek the solution of ${\mathbf A}$ in the form of a wave propagating along $\hat{\mathbf z}$ with transverse polarization (say along $\hat{\mathbf x}$). Such a solution takes the form:
\begin{equation} 
{\mathbf A} = [a(x,y,z)e^{ikz} + c.c]\hat{\mathbf x}, 
\label{eq:5}
\end{equation}   
where $k=k_0\sqrt{\epsilon}$ and $a(x,y,z)$ is the slowly varying (with respect to z) amplitude. Using equations (\ref{eq:3}) and (\ref{eq:5}), $\nabla\phi$ can be evaluated to be
\begin{equation} 
\nabla\phi = (ik_{0}\epsilon)^{-1} \nabla{\left(\frac{\partial a}{\partial x} \exp{(ikz)}\right)}, 
\label{eq:6} 
\end{equation} 
which subsequently leads to the expression of ${\mathbf E}$ as follows.
\begin{equation} 
{\mathbf E} = ik_{0} \left(a\hat{\mathbf x} + \frac{i}{k} \frac{\partial a}{\partial x} \hat{\mathbf z} \right) \exp{(ikz)}. 
\label{eq:7}
\end{equation} 
It is clear from equation (\ref{eq:7}) that the paraxial approximation inevitably leads to a longitudinal component. Substituting expression (\ref{eq:5}) in equation (\ref{eq:4}) and making use of the slowly varying approximation one can write down the paraxial equation for a light beam:
\begin{equation} 
\frac{\partial^{2} a}{\partial x^{2}} + \frac{\partial^{2} a}{\partial y^{2}} + 2ik \frac{\partial a}{\partial z} = 0. 
\label{eq:8}
\end{equation} 
The solution of (\ref{eq:8}) can be expressed as:
\begin{equation} 
a(x,y,z) = \int {\widetilde{a}(q_{x},q_{y})\exp{[i(q_{x}x + q_{y}y)]}\exp{\left(-i \left(\frac{{q_{x}}^{2} + {q_{y}}^{2}}{2k^{2}}\right)z\right)}dq_{x}dq_{y}}.
\label{eq:9}
\end{equation} 
For localized profiles of the form $\widetilde{a}(q_{x},q_{y})$, one has solutions in the form of a beam. For example, for
\begin{equation} 
\widetilde{a}(q_{x},q_{y}) = \pi{W_{0}}^{2} \exp{\left(-{W_{0}}^{2}\left(\frac{{q_{x}}^{2}+{q_{y}}^{2}}{4}\right)\right)}, 
\label{eq:10}
\end{equation} 
one recovers the fundamental gaussian beam:
\begin{equation} 
a(x,y,z) = \frac{W_{0}}{W(z)}\exp{\left(\frac{-r^{2}}{W^{2}(z)}\right)}\exp{\left(\frac{ikr^{2}}{2R(z)}\right)}\exp{\left(-i\varphi(z)\right)}\exp(ikz). 
\label{eq:11}
\end{equation}     
The expressions for the beam waist $W(z)$, radius of curvature $R(z)$ and Gouy phase $\varphi(z)$ in the above expression are standard. Following a similar route, the expression for the general Hermite-Gaussian beam can be calculated. In order to calculate the reflected beam profiles for higher order Hermite-Gaussian modes, we have taken the incident beam amplitude as follows:
\begin{equation}
\begin{array}{rcl}
a_{m,n} &=& \frac{W_{0}}{W(z)}\exp{\left(\frac{-r^{2}}{W^{2}(z)}\right)}\exp{\left(\frac{ikr^{2}}{2R(z)}\right)}\exp{\left(-i\varphi(z)_{m,n}\right)} \\
          & &\times\exp(ikz)H_{m}\left(\frac{\sqrt{2}x}{W(z)}\right)H_{n}\left(\frac{\sqrt{2}y}{W(z)}\right), 
\end{array}
\label{eq:19}
\end{equation}
where $W(z)$ and $R(z)$ have the same meaning as before, but the Guoy phase takes on the definition:
\begin{equation} 
{\varphi}_{m,n}(z) = (m+n+1)\tan^{-1}\left(\frac{z}{z_{r}}\right),  
\label{eq:20}
\end{equation}
where $z_{r} = \frac {k{w_{0}}^{2}}{2}$. 
\par
In the context of the problem of calculating the reflection coefficient, we now specify the geometry. The geometry is akin to that considered by Aiello and Woerdman \cite{Woerdman}. We consider an interface (at $z = 0$) between vacuum and a dielectric medium with dielectric constant $\epsilon$. Let the beam be incident at an angle $\theta$ to $\hat{z}$ and with vector potential polarized along $\hat{y}$. Let also $z'$ axis coincide with the direction of propagation of the beam. The primed and unprimed axes are oriented at $\theta$ to each other.  The relation between the unprimed and primed co-ordinates are given by
\begin{equation} 
x' = x\cos\theta - z\sin\theta,~~ y' = y,~~ z' = z\cos\theta +x\sin\theta. 
\label{eq:12}
\end{equation} 
For a fundamental gaussian beam of the form given by (\ref{eq:10}), ${\mathbf E}$ reads as follows.
\begin{eqnarray}
{\mathbf E} &=& \int 
\left(\hat{\mathbf y'} - \frac{q'_y}{k}\hat{\mathbf  z'}\right)
\tilde{a}(q'_{x},q'_{y})\nonumber\\
&& ~~ \times\exp{\left[i\left(q'_{x}x' + q'_{y}y'- \frac{{q'_{x}}^{2} + {q'_{y}}^{2}}{2k^{2}}z'+kz'\right)\right]}d^{2}q'.
\label{eq:13}
\end{eqnarray} 
This expression is consistent with the result of Chen {\it et al} \cite{Carlchen}, that the choice of a zero transverse field component at the input boundary plane ensures the vanishing of the component throughout the propagation in the same medium. The consistency of the magnitude of the total wave vector of every spatial component under the paraxial approximation $ \sqrt{{q_{x}}^{2} + {q_{y}}^{2}} << k $ can easily be verified.
\begin{equation} 
{q'_{x}}^{2} + {q'_{y}}^{2} + k^{2}(1 - \frac{{q'_{x}}^{2} + {q'_{y}}^{2}}{2k^{2}})^{2} \approx k^{2}.
\label{eq:14}
\end{equation} 
Note also that under very weak transverse variation of the field when the inequality $\frac{q'_y}{k} (<< 1)$ holds, (\ref{eq:13}) leads to a simple s-polarized beam all throughout \cite{Hiroshi}. Henceforth, we refer to this approximation as the strong paraxial approximation. 
\par Moving to the unprimed coordinate system one can write down a typical plane wave component of ${\mathbf E}$ as:
\begin{equation} 
({e_{i}}^x\hat{\mathbf x} + {e_{i}}^y\hat{\mathbf y} + {e_{i}}^{z}\hat{\mathbf z})\widetilde{a}(q_{x},q_{y})\exp{[i(q_{x}x + q_{y}y + q_{z}z)]},
\label{eq:15}
\end{equation} 
where ${e_{i}}^x = \frac{-q'_{y}}{k}\sin\theta$, ${e_{i}}^y = 1 $, ${e_i}^{z} = \frac{-q'_{y}}{k}\cos\theta$ and $q_{x}$, $q_{y}$ and $q_{z}$ are given by:
\begin{eqnarray} 
q_{x} &=& q'_x\cos\theta + k\sin\theta\left(1 - \frac{{q'_{x}}^{2}+{q'_{y}}^{2}}{2k^{2}}\right), \nonumber \\
q_{y} &=& q'_y \label{eq:16}, \\
q_{z} &=& q'_x\sin\theta - k\cos\theta\left(1 - \frac{{q'_{x}}^{2}+{q'_{y}}^{2}}{2k^{2}}\right). \nonumber
\end{eqnarray}
It is again easy to check the consistency of the total wave vector for a spatial component under the paraxial approximation using above equations 
\begin{equation} 
{q_{x}}^{2} + {q_{y}}^{2} + {q_{z}}^{2} \approx k^{2}.
\label{eq:17}
\end{equation} 
\par
In what follows, we neglect the terms proportional to $\frac{q}{k}$ in the amplitude, while retaining the full dependence in the phase factors. Thus the beam reduces to an s-polarized beam with only non-vanishing component along $\hat{\mathbf{y}}$. Then, one can easily calculate the reflected component by multiplying the amplitude of the spectral component with the reflection coefficient $r_{s}$ which is given by
\begin{equation} 
r_{s} = \frac{1 - \frac{q_{z_{0}}}{q_z}}{1+ \frac{q_{z_{0}}}{q_z}}, 
\label{eq:18} 
\end{equation}
where $q_{z_{0}} = \sqrt{k_0^{2} - ({q_{x}}^{2} + {q_{y}}^{2})}$. In order to validate the above approximation, one has to have a broad beam (close to plane wave profile), so that $ \frac{{q'_{x}}^{2} + {q'_{y}}^{2}}{k^{2}}<<1 $. The broader the beam, the smaller is the GH effect. 
Note that plane waves do not exhibit any GH shift. Thus one has to exercise care in order to choose the parameter domain where the shift, albeit small, is still observable, if one wants to describe the phenomenon with a model of an incident gaussian beam with a given polarization. For a tightly focused beam, the paraxial approximation forces one to include the longitudinal component. With the longitudinal component present, the calculation of the reflection and transmission coefficients poses a formidable problem.
\section{Numerical results and plots}
In this section we follow the procedure detailed in the previous section to calculate the beam profile on the interface before and after the reflection. To be specific, we take a beam described by a scalar function with s-polarization. We find the 2-d Fourier components incorporating the oblique incidence. Each Fourier component is multiplied by the corresponding reflection coefficient. Then an inverse Fourier transform leads to the reflected beam profile at $z=0$. We have considered the cases of fundamental and higher order Gaussian beams. We show that identity of the beam is retained in the case of Gaussian-Hermite beams.
 In all our calculations we have taken $\sqrt{\epsilon} = 1.7321$ (the medium of incidence), $\epsilon_{0} = 1.0$, $kW_{0} = 40$, unless specified otherwise. The incident beam is assumed to be focused at $z= 0$ (at the interface). 
\par
While choosing the relevant parameters, special attention was given to two aspects. First, the sampling frequency was taken to be more than the Nyquist rate \cite{Goodman}. Second, the parameter domain was chosen such that the strong paraxial approximation was valid. For an angle of incidence of $\theta = 40^{\circ}$ and $kW_{0} = 40$, for the $y$ direction, we take the sampling interval to be $\Delta y/W_0  = 0.1020$. Sampling is done inside a square area of side of $15$ units. In the inverse Fourier space, the sampling interval is $\Delta q_{y} = 0.0602$. Let $\widetilde{q_{y}}$ represent the value at which the amplitude falls off, say, to $10^{-5}$ times the peak value. The value of  $\widetilde{q_{y}}$ gives an estimate of the angular dispersion of the incident beam. For a fundamental Gaussian beam with $kW_{0}=40$, we have $\widetilde{q_{y}} \approx 0.17$. Thus the chosen parameters alleviate both the above concerns.
\begin{figure}[htbp]
\centering
\includegraphics[width=10cm]{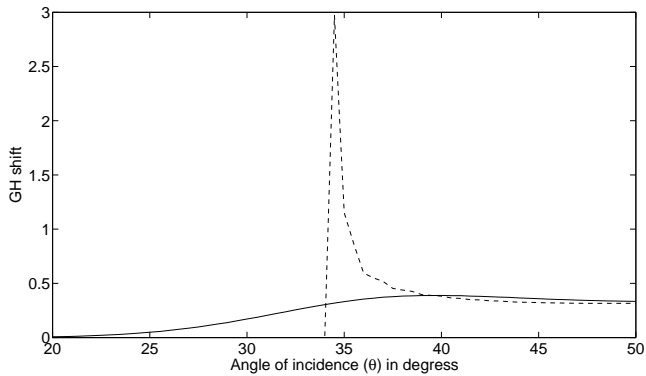}
\caption{Normalized GH shift as a function of angle of incidence $\theta$. Solid line: numerically computed GH shift; dashed line: shift predicted from the Artmann formula}
\label{fig1}
\end{figure}
\par
We first use the numerical method to predict the GH shift for a two dimensional Gaussian beam. Figure \ref{fig1} shows the dependence of the GH shift (normalized to the beam waist $W_0$) on the angle of incidence $\theta$ for a 2-d fundamental Gaussian beam. The GH shift values computed using the procedure described in the previous section (solid line) are contrasted with those predicted by the Artmann formula \cite{McGuirk} (dashed line). It is clear from figure \ref{fig1} that for angles larger than the critical angle and away from it, the GH shift computed numerically falls off and compares well with that predicted by the Artmann formula. Near the critical angle, it is well known that the Artmann formula has a singularity \cite{Lai}. Thus the stationary phase approximation, which is used to derive the Artmann formula, is not applicable near the critical angle. Our numerical results are consistent with the fact that the GH shift must show a smooth, continuous and finite variation near the critical angle \cite{Lai}. 
\par
\begin{figure}[htbp]
\centering
\includegraphics[width=15cm]{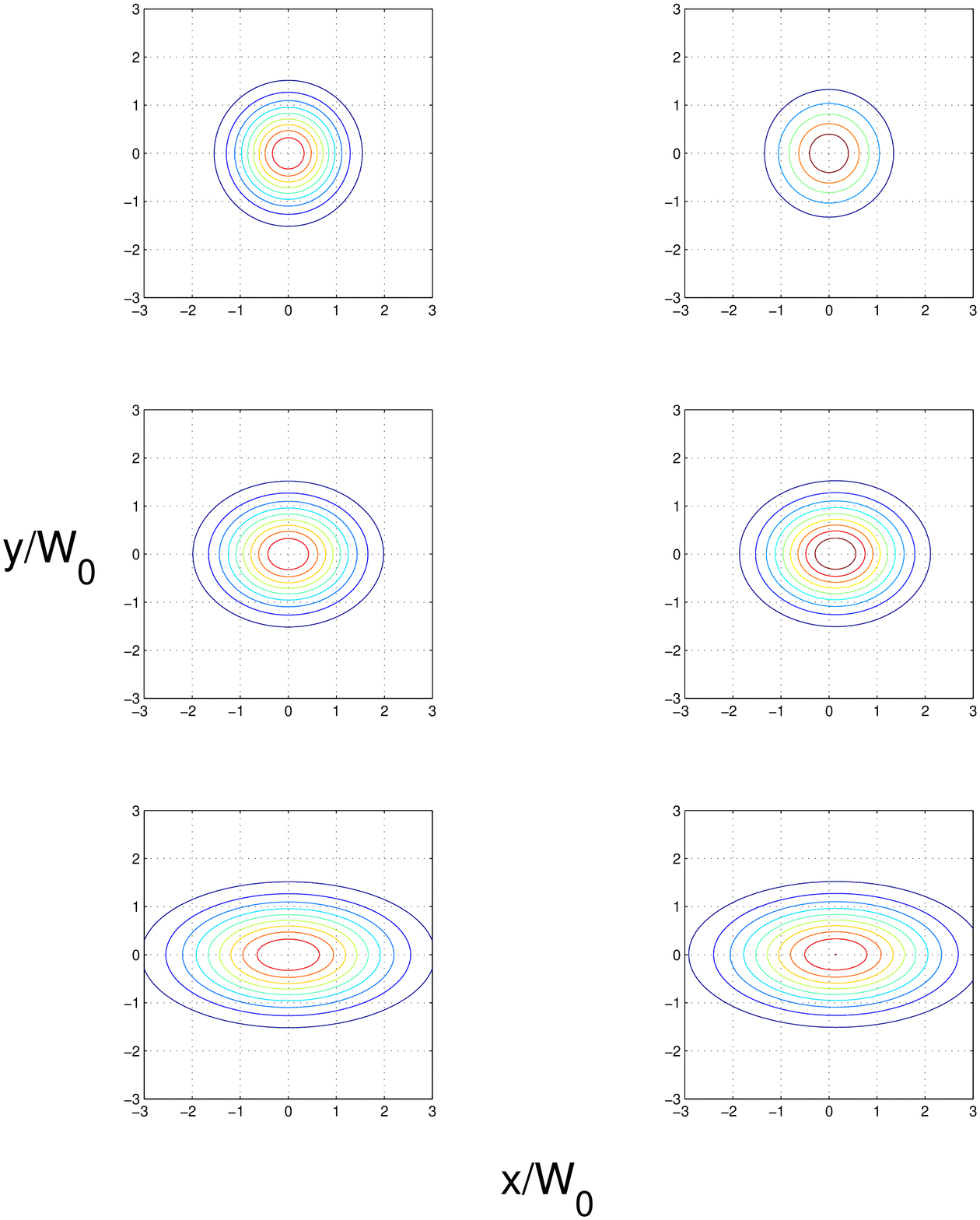}
\caption{Incident (left) and reflected (right) profiles for fundamental Gaussian beams. The angles of incidence are top: $10^{\circ}$, middle: $40^{\circ}$, bottom: $60^{\circ}$}
\label{fig2}
\end{figure}

Having validated our numerical procedure in the two dimensional case, we now use it to analyze three dimensional Gaussian beams. The profiles on the interface for the incident and reflected fundamental Gaussian beam are shown in figure \ref{fig2}. The left (right) pane in each horizontal line shows the contour of the incident (reflected) beam at the interface. Horizontal (vertical) axis is the normalized $x (y)$ direction. It 
is clear from figure \ref{fig2} that for small and large angles of incidence, there is practically no shift of the fundamental Gaussian beam. Close to the critical angle $\theta_{cr}$ 
\begin{figure}[htbp]
\centering
\includegraphics[width=15cm]{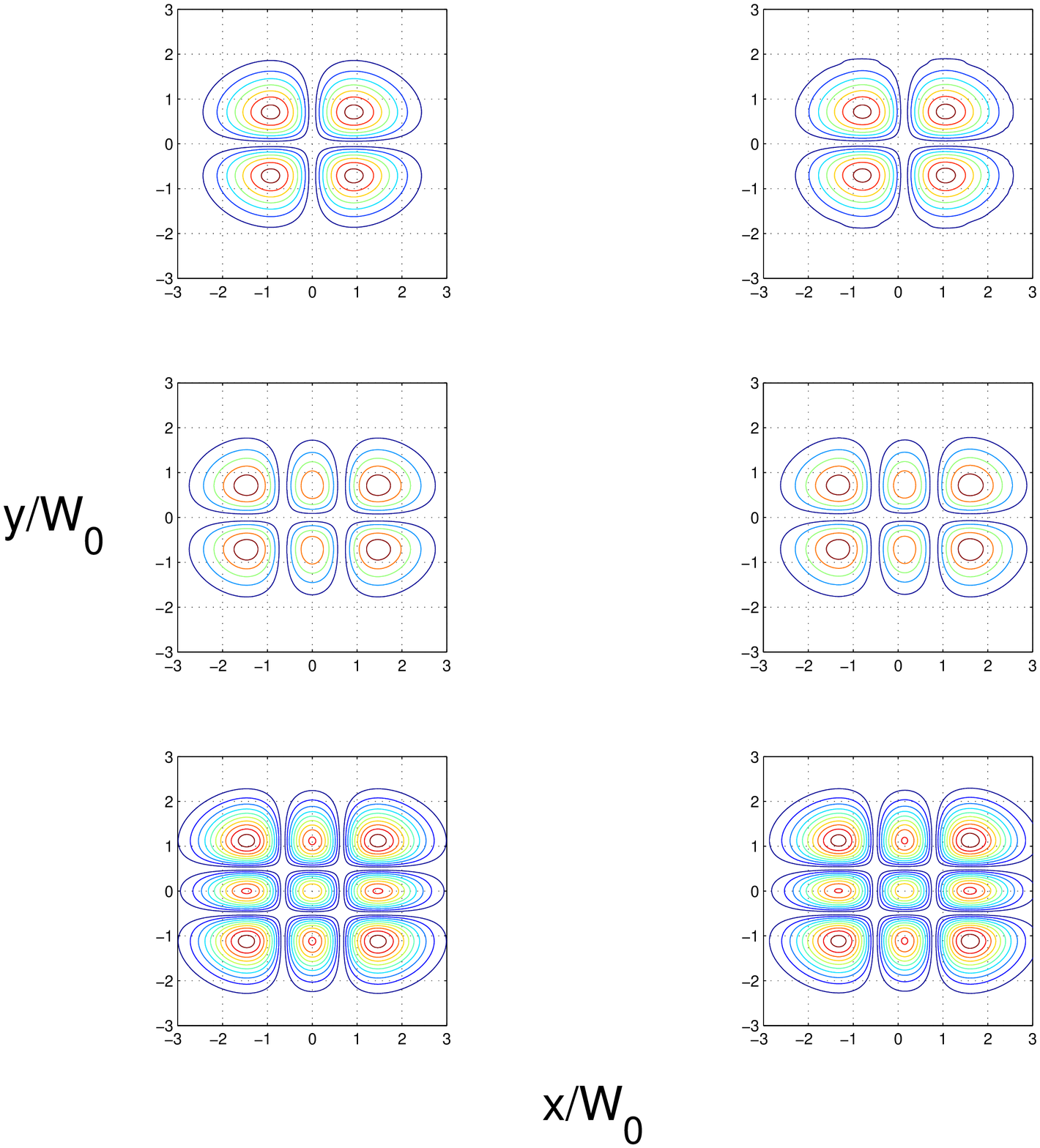}
\caption{Incident (left) and reflected (right) beam profiles for $TEM_{11}$ (top), $TEM_{21}$ (middle), $TEM_{22}$ (bottom), incident at $40^{\circ}$.}
\label{fig3}
\end{figure}
($= \arcsin(\sqrt{\epsilon_0/\epsilon})$), the shift is discernible. Slightly above the critical angle, the shift achieves the maximum value. 
\par
Figure \ref{fig3} shows the results for $TEM_{11}$ beam incident at $40^{\circ}$ (approximately where the GH shift is largest). It is clear from Figure \ref{fig3} that even after reflection the beam identity of the $TEM$ beam is intact, though there is a slight difference in the heights of the pair of lobes. The shift becomes negligible for small and large values of the angle of incidence.  Figure \ref{fig3} also shows the incident and reflected profiles for $TEM_{21}$ (middle) and $TEM_{22}$ (bottom) cases, respectively. Even for For Hermite-Gaussian beams the GH shift is small for small and large angles of incidence, in this case peaking at around $40^{\circ}$. The GH shift for near normal incidence is almost negligible.
\par
We now focus our attention to beam distortion under reflection near the critical angle. For loosely focussed beams, it was shown that the distortion is negligible for Hermite-Gaussian beams. This is not the case for Laguerre-Gaussian beams because of the angular momentum carried by the beam. The inherent orbital angular momentum \cite{Allen} of the beam and the lateral Feodorov-Imbert shift are responsible for the distortion of the beam profile upon reflection \cite{Hiroshi}. Thus, loosely focused higher order Hermite-Gaussian beams can retain their identity upon reflection, while the Laguerre-Gaussian beams can be distorted beyond recognition.
\begin{figure}[htbp]
\centering
\includegraphics[width=10cm]{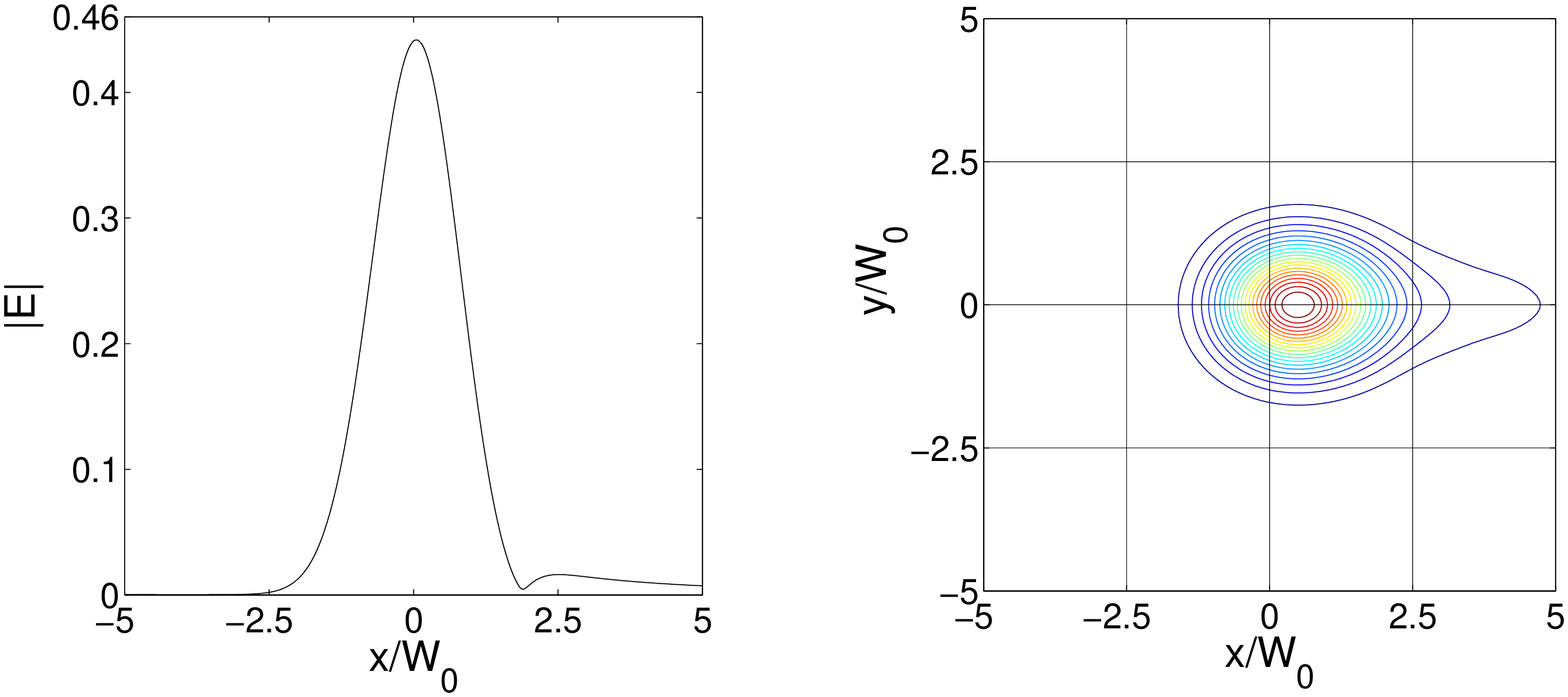}
\caption{The reflected beam profile for 2-d (left) and 3-d (right) Gaussian beam incident at $25^{\circ}$.}
\label{fig4}
\end{figure}
\par
Until now, we have been looking at broad (loosely focused) beams which fall well within the realm of validity of the paraxial approximation. We now look at tightly focused beams which fall at the periphery of this regime. We qualitatively describe the effects of reflection on a tight Gaussian beam. Figure \ref{fig4} shows the incident and reflected profiles for a tightly focused ($kW_{0}=15$) fundamental beam for 2-d and 3-d simulations (left and right panels, respectively, in the figure). The reflected profile shows a splitting in the beam, with the newly developed lobe having very little intensity compared to the main lobe. This observation is in step with other investigations of a similar kind where beam splitting upon reflection has been reported \cite{Lin,Yuhang}. It has to be kept in mind that very tight beams, by default, fall in the regime of non-paraxial beams and that in this regime the longitudinal component cannot be summarily neglected.
\section{Conclusion}
In conclusion, we studied the reflection of Gaussian beams from a dielectric-vacuum interface and showed that for tight beams it is advisable to take into account the longitudinal component of the field under paraxial approximation, while one can use standard scalar beams with some given polarization in the opposite case. We numerically studied the reflected beam profiles and demonstrated the GH shift for fundamental and higher order Gaussian beams. Calculations were carried out after neglecting the longitudinal component of the electric field, which is justified in the strong paraxial approximation regime. We also concluded that higher order Hermite-Gaussian beams retain their identity after reflection unlike Laguerre-Gaussian beams.  For tight beams it is indicative that beam splitting occurs upon reflection, but for more precise results the longitudinal component must be retained in this case. Further investigation is needed to tackle the problem of reflection and associated GH shift of vector beams without making the strong paraxial approximation. Work is underway on enhancing the GH shift of higher order Gaussian beams using various layered media with metal-dielectric structures and metamaterials.
\section*{Acknowledgements}
One of the authors (SDG) is thankful to Girish Agarwal for many helpful discussions. The authors are also thankful to Dr. Subimal Deb for discussions and help with the manuscript. This work was supported by a Project  from  the Department of Science and Technology, Govt. of India. 

\end{document}